\documentstyle[twoside,fleqn,espcrc2,epsfig]{article}

\newcommand{\AmS}{{\protect\the\textfont2
  A\kern-.1667em\lower.5ex\hbox{M}\kern-.125emS}}

\title{
Kondo Quartet}
\author{A.Georges\address{
        Laboratoire de Physique Th\'{e}orique de l'Ecole
Normale Sup\'{e}rieure \\ 
        24, rue Lhomond, 75231 Paris Cedex 05, France}%
        \thanks{
Unit\'e propre du CNRS (UP701) associ\'ee \`a l'ENS et \`a 
l'Universit\'e Paris-Sud}
        and 
        A.Sengupta 
\address{Raman Research Institute\\
         C. V. Raman Avenue,
        Bangalore 560080, India}%
\thanks{Also at: Serin Physics Laboratory, Rutgers University, 
Piscataway, NJ 08854, USA }} 
       
\begin{document}

\begin{abstract}
This article describes some recently obtained results
on the low-energy properties of the "Kondo quartet" model of 
two spin-1/2 impurities interacting with two
channels (flavours) of conduction electrons. 
We shall particularly emphasize the
connections between conformal field-theory methods and bosonisation approaches,
which are first illustrated on the example of
the single-impurity, two-channel Kondo problem. 
This article is dedicated to the memory of Claude Itzykson, and will 
appear in the Proceedings of the Conference "Advanced Quantum Field 
Theory", held in La Londe Les Maures in Sept, 1996 (Nucl. Phys. B, 
Proc. Supp., V.Rittenberg, J.Fr\"{o}lich and A.Schwimmer eds.). 
\end{abstract}

\maketitle

\section{The two channel Kondo model}
\label{2channel}

Two methods leading to a low-energy solution of the single impurity, 
two channel Kondo model \cite{NB}  have been recently put forward. 
One, due to 
Affleck and Ludwig \cite{AL,ALPC}, uses the 
general framework of (boundary) conformal field theory (CFT). 
The other, due to 
Emery and Kivelson \cite{EK,EKetc}, relies 
on abelian bosonisation and establishes an exact  
mapping onto a resonant-level model 
which reduces to free fermions for a special value of the Kondo 
coupling $J_{z}$. This is analogous to the Toulouse limit \cite{GT} 
of the single-channel case. The aim of this section is to provide the 
reader with an introduction to the two channel Kondo physics by 
reviewing partly these methods  
and, more importantly, by establishing the precise connection between 
the two approaches. This will prove to be very useful in the study of the 
corresponding two impurity 
problem.

\subsection{Model and Symmetries} 

We formulate the model in terms of left-moving chiral fermions
$\psi_{i\alpha} (x)$ on the
full axis
$-\infty<x<+\infty$, and of a spin-$1/2$ impurity spin $\vec{S}$ placed 
at the origin. $i=1,2$ is a channel index, and
$\alpha=\uparrow,\downarrow$ is a spin index. 
The hamiltonian is written as:
\begin{eqnarray} 
&H = i v_F \sum_{i\alpha} \int_{-\infty}^{+\infty} dx
\psi_{i\alpha}^{\dagger}(x){{\partial}\over{\partial x}} 
\psi_{i\alpha}(x)\\
&+J_z S_z {\cal J}_z(0) + J (S_x {\cal J}_x(0)+S_y {\cal J}_y(0)) 
\end{eqnarray}
In these formula, 
$\vec{{\cal J}}(x)=
\sum_{i=1}^{2}\sum_{\alpha\beta}
\psi_{i\alpha}^{\dagger}(x) {{\vec{\sigma}_{\alpha\beta}}\over{2}}
\psi_{i\beta}(x)$ 
denotes the total conduction 
electron spin current at 
position $x$. 
A Kondo coupling which is anisotropic in spin space has been considered, 
for reasons which will become clear below. 
The anisotropy $J-J_z$ is actually  
{\it irrelevant} at the non-trivial fixed point. As demonstrated by 
Nozieres and Blandin \cite{NB} and widely discussed since, 
both the free fermion ($J=0$) and the strong-coupling ($J=+\infty$) fixed 
points are unstable under renormalization. The low-energy physics is 
controlled by an intermediate coupling fixed point, which has non Fermi 
liquid properties $\chi_{imp}\,\sim\,C/T\,\sim\,\mbox{ln}1/T$. 

The interacting 
hamiltonian has, in the isotropic case $J_z=J$, a global symmetry 
$SU(2)_{spin}\otimes SU(2)_{flavour}\otimes U(1)_{charge}$. 
This global symmetry group is {\it smaller} than the 
symmetry of the non-interacting theory with $J_z=J=0$, which has 
a $U(4)=SU(4)\otimes U(1)$ invariance mixing all four species $i,\alpha$. 
This invariance is partially broken by the Kondo interaction. 
In fact, as explained in detail in \cite{ALPC}, the non-interacting 
theory has a full $SO(8)$ global symmetry 
including transformations
which mix flavour and charge, and the 
interacting theory itself has a larger global symmetry than the 
spin-flavour-charge one, namely an  
$SU(2)_{spin}\otimes SO(5)_{cf}$ invariance.
We shall make use of this larger symmetry later in this paper. 

The basic assumption of the CFT approach is that, at a fixed point, 
these global symmetries are promoted to local conformal symmetries. The 
symmetry algebra becomes the Kac-Moody algebra built on the global 
symmetry group of the interacting theory, namely:
\begin{equation} 
\widehat{SU}_2(2)_{s}\otimes \widehat{SU}_2(2)_{f}\otimes
\widehat{U}(1)_{c}
\end{equation}
Here, $\widehat{SU}_k(2)$ stands for the level-$k$
$SU(2)$ Kac-Moody algebra \cite{PG}. 
The Kac-Moody symmetry 
puts strong constraints on the theory. 
In particular, it dictates the general form of the finite-size 
spectrum {\it at any fixed point of the model} as:
\begin{equation}
{{L\Delta E}\over{\pi v_F}} = {{j(j+1)}\over{4}}+m\,+\,
{{j_f(j_f+1)}\over{4}}+m_f\,+\,{{Q^2}\over{8}}+m_c
\label{fsscft}
\end{equation}
In this expression, $\Delta E=E-E_{gs}$ are the excitation energies, 
$v_F$ is the Fermi velocity and $L$ the {\it radial} length of the system. 
The quantum number $j$ labels the representation of the
$\widehat{SU_2(2)}_{spin}$ algebra and takes the possible values $j=0,1/2,1$.
Similarly, $j_f=0,1/2,1$ labels the representation of
$\widehat{SU_2(2)}_{flavour}$ and $Q$ is a (positive or negative) integer
corresponding to the charge of $\widehat{U(1)}_c$. $m,m_f,m_c=0,1,2,3,...$'s
are integers corresponding to towers of states built on these primary
states. The degeneracies are also specified uniquely in terms of
these quantum numbers.

The general form (\ref{fsscft}) dictates the {\it possible} energy levels,
but does not specify which set of quantum numbers $j,j_f,Q$ {\it actually
appears} in the spectrum at a given fixed point. The strategy followed
by Affleck and Ludwig is first to identify the appropriate selection
rules for the free fermion fixed point ($J=0$). For this fixed point,
all energy
levels can be built following the Pauli principle. 
A major insight is then that the spectrum at the
non-trivial infra-red stable fixed point can be obtained 
from a 'fusion principle'. 
Specifically, the spectrum is obtained by acting on the primary operator 
in the spin sector associated with a given state of the 
free-fermion
fixed point with the operator $(j=1/2)$ of the
$\widehat{SU_2(2)}_{spin}$ algebra (leaving $j_f$ and $Q$ unchanged).
Each admissible values of $j=j_{FF}$ at the non-interacting fixed point then
gives rise to at most two admissible values at the new fixed point,
according to the fusion rule of CFT:
\begin{eqnarray}
\nonumber &j=0\rightarrow j'=1/2\\
\nonumber &j=1/2\rightarrow j'=0\, \mbox{and} \, j'=1\\
&j=1\rightarrow j'=1/2
\label{fusion}
\end{eqnarray}
Physically, this fusion principle reflects the (partial) screening of the 
impurity spin by the conduction electrons. 
The conformal towers appearing in the resulting spectrum 
are detailed in Ref.\cite{AL,ALPC}. 
This construction can be extended to find     
the operator content at the 
non-trivial fixed point (by a double fusion).

\subsection{Bosonisation}

The decomposition of a fermion $\psi_{i\alpha}(x)$ into 
spin-flavour-charge degrees of freedom can be made explicit in two 
possible ways. The first is to make use of `non-abelian bosonisation' 
which has the advantage of respecting all global symmetries explicitly. 
This route is the natural one within the CFT approach. The fermion field 
is written as:
\begin{equation}
\psi_{i\alpha}(x)^{+} = g^{+}_{\alpha}\,h^{+}_{i}\,e^{i\Phi_c/2}
\label{nonab}
\end{equation}
In this expression, $g^{+}_{\alpha},\,h^{+}_{i}$ are Wess-Zumino-Witten fields  
belonging to the fundamental representation of the $\widehat{SU}_2(2)_{s}$ 
and $\widehat{SU}_2(2)_{f}$ spin and flavour algebra, respectively, and 
$\Phi_c$ is a charge boson. Alternatively, one may take a 
more `pedestrian' (and often more explicit) route by using  
abelian bosonisation, even though 
the global $SU(2)$ invariances are no longer manifest. This is the 
starting point of the approach due to Emery and Kivelson \cite{EK}, who introduce 
four boson fields such that:
\begin{equation}
\psi_{i\alpha}(x)={{1}\over{\sqrt{2\pi a_0}}} e^{-i\Phi_{i\alpha}(x)}
\end{equation}
where:
\begin{equation}
\Phi_{i\alpha}(x)=\sqrt{\pi}\{\int_{-\infty}^{x}dx' \Pi_{i\alpha}(x') -
\phi_{i\alpha}(x) \}
\end{equation}
with:
\begin{equation}
[\phi_{i\alpha}(x) , \Pi_{j\beta}(x')] = i \delta_{ij} \delta_{\alpha \beta}
\delta(x-x')
\end{equation}
Our conventions for boson fields  
are such that an operator $O=e^{i g\Phi}$ has dimension 
$\Delta=g^2/2$ ({\it i.e} its correlation function behaves as 
$<O(0)O(x)>\sim 1/x^{2\Delta}$ for $|x|\rightarrow +\infty$).

Linear combinations are then formed, corresponding to 
charge $\phi_c$, spin $\phi_s$, flavor $\phi_f$ and
spin-flavor $\phi_{sf}$ degrees of freedom (similarly
$\Phi_c,\Phi_s,\Phi_f,\Phi_{sf}$):
\begin{eqnarray}
\phi_c = {{1}\over{2}} (\phi_{1\uparrow}+\phi_{1\downarrow}+
\phi_{2\uparrow}+\phi_{2\downarrow})\\
\phi_s = {{1}\over{2}} (\phi_{1\uparrow}-\phi_{1\downarrow}+
\phi_{2\uparrow}-\phi_{2\downarrow})\\
\phi_f = {{1}\over{2}} (\phi_{1\uparrow}+\phi_{1\downarrow}-
\phi_{2\uparrow}-\phi_{2\downarrow})\\
\phi_{sf} = {{1}\over{2}} (\phi_{1\uparrow}-\phi_{1\downarrow}-
\phi_{2\uparrow}+\phi_{2\downarrow})
\end{eqnarray}
In term of these fields, the components of the total spin current 
read:
\begin{eqnarray}
\nonumber
&{\cal J}_x = {{1}\over{\pi a_0}} cos\Phi_s cos\Phi_{sf}\,\,,\,\,
{\cal J}_y = {{1}\over{\pi a_0}} sin\Phi_s cos\Phi_{sf}\\
&{\cal J}_z = - {{1}\over{2\pi}} {{\partial\Phi_s}\over{\partial x}}
\end{eqnarray}
Only $\Phi_s$ and $\Phi_{sf}$ enter these expressions. This is 
expected since these fields correspond to the sum and the difference 
of the bosons associated with spin degrees of freedom for each channel, 
namely $\Phi_{1\uparrow}-\Phi_{1\downarrow}$ and 
$\Phi_{2\uparrow}-\Phi_{2\downarrow}$. 
Furthermore, note that among the four independent real combinations 
of these fields, only {\it three} appear: $sin\Phi_{sf}$ is not 
involved in the total spin current. 
That three real fields enter the current reflects the fact that 
the $\widehat{SU}_2(2)$ Kac-Moody algebra has central charge $c=3/2$. 
These fields have dimension $1/2$ and can alternatively 
be considered as three {\it real fermionic fields} (Majorana fermions) 
$\chi^a\,\,,\,\,a=x,y,z$:
\begin{equation}
\chi^x=sin\Phi_s\,\,,\,\,
\chi^y=cos\Phi_s\,\,,\,\,
\chi^z=cos\Phi_{sf}
\label{majoranas}
\end{equation}
in terms of which the total spin current reads:
\begin{equation}
{\cal J}^a = i \epsilon_{abc} \chi^b \chi^c
\label{majocurrent}
\end{equation}
This representation of the $\widehat{SU}_2(2)$ algebra in terms of 
three Majorana fermions is well known in CFT \cite{PG}, 
and will prove to be most 
useful below in the solution of the two impurity problem. 

The hamiltonian is easily written in terms of the boson fields. 
The free (kinetic energy) part reads:
\begin{equation}
H_0 = {{v_F}\over{4\pi}} \sum_{l=s,c,f,sf} \int dx  
({{\partial \Phi_l}\over{\partial x}})^2 
\end{equation}
while the Kondo interaction takes the form:
\begin{eqnarray}
&H_{int}= -{{J_z}\over{2\pi}} S^z {{\partial\Phi_s}\over{\partial x}}(0)+\\
&{{J}\over{\pi a_0}} [S^x cos\Phi_s(0)+S^y sin\Phi_s(0)] cos\Phi_{sf}(0)
\end{eqnarray}

\subsection{Fixed-point boundary condition:  
the fusion rule hypothesis derived from the bosonisation approach}

In order to proceed with the analysis of $H_{int}$, 
Emery and Kivelson \cite{EK} 
observe that $\Phi_s$ can 
be eliminated from the terms involving $S^x$ and $S^y$,  
by performing a rotation of the fields along the $S^z$ direction, 
{\it i.e} acting with the operator:
\begin{equation}
{\cal U} = \exp(i S^z \Phi_s(0))
\label{ektrans}
\end{equation}
The transformed hamiltonian reads:
\begin{eqnarray}
&\widetilde{H}_0\equiv {\cal U} H_0 {\cal U}^{+}
= H_0 +v_F S^z {{\partial\Phi_s}\over{\partial x}}(0) \\
\nonumber &\widetilde{H}_{int}\equiv {\cal U} H_1 {\cal U}^{+}
= -{{J_z}\over{2\pi}} S^z{{\partial\Phi_s}\over
{\partial x}}(0) + {{J}\over{\pi a_0}} S^x cos\Phi_{sf}(0)
\end{eqnarray}
Hence, for the special value of the coupling, $J_z=2\pi v_F$, 
$\Phi_s$ no longer enters the hamiltonian, which takes the 
very simple form:
\begin{equation}
H_0+ {{J}\over{\pi a_0}}\, S^x cos\Phi_{sf}(0)
\end{equation} 
This describes a resonant level model for the 
Majorana fermion $\chi^z=cos\Phi_{sf}$, with scattering 
at the Fermi level. So $\chi^z$ (which was left unchanged by 
the rotation ${\cal U}$)  
suffers a $\pi/2$ phase shift or, more appropriately for a real 
object, suffers a change of boundary condition from periodic 
($\chi^z(0^+)=\chi^z(0^-)$) to antiperiodic ($\chi^z(0^+)=-\chi^z(0^-)$). 
The 'twist' operator $\sigma_z$ which achieves this change of boundary  
conditions when acting on $\chi^z$  
has dimension $\Delta(\sigma_z)=1/16$, corresponding to 
the shift in the ground-state energy under a change of boundary 
conditions of one free Majorana fermion: 
$L\Delta E/\pi v_F=1/16$. Hence $\Delta(\sigma_z)=1/16$.
 
In contrast, the 
{\it transformed} fields $\widetilde{\chi}^x$ and 
$\widetilde{\chi}^y$ drop 
out from the hamiltonian at the solvable point, and hence remain 
free and unaffected. The relation of these transformed fields 
to the original ones is easily worked out by considering the 
Dirac fermion:
\begin{equation}
\psi_{s}\equiv {{1}\over{\sqrt{2\pi a_0}}} e^{-i\Phi_{s}}
\end{equation}
and observing that:
\begin{eqnarray}
&\nonumber\widetilde{\psi_s}(x) \equiv {\cal U} \psi_s(x) {\cal U}^{+}
= e^{iS^z\Phi_s(0)} {{e^{-i\Phi_s(x)}\over{\sqrt{2\pi a_0}}}} 
e^{-iS^z \Phi_s(0)}\\
& = e^{-i\pi S^z sign(x)} \psi_s(x)
\end{eqnarray}
Since $S^z$ has eigenvalues $\pm 1/2$, one sees from this equation 
that $\psi_s(0^+)=e^{\pm i\pi}\psi_s(0^-)$ and hence that 
the {\it original} field $\psi_s$ also suffers a  
phase shift $\delta=\pi/2$ at the solvable point. 
In this case, the operator $\cal{U}$ which achieves this has 
the simple boson representation (\ref{ektrans}) because it 
changes both components 
of a Dirac fermion. Its dimension is thus 
$1/8=\delta^2/2\pi^2$, twice that of a twist operator for a single Majorana 
fermion.

Hence, the crucial feature of the Emery-Kivelson solvable point is that 
the three (original) Majorana fermions $\chi^{x,y,z}$ introduced 
in Eq.(\ref{majoranas}) in order  
to represent the total spin current suffer a {\it change of 
boundary condition from periodic to antiperiodic}. 
(Actually, as discussed below, we need to consider both the antiperiodic 
and periodic Majorana sectors to construct the free fermion spectrum of 
the four original Dirac fermions with e.g antiperiodic boundary 
conditions. At the Kondo fixed point, the Majoranas which were originally 
periodic become antiperiodic, and the Majoranas which were originally
antiperiodic formally become periodic, but in fact do not contribute to 
the spectrum at the Kondo fixed point.)    
This interpretation of the boundary condition associated with the non-trivial 
fixed point has also been discussed independently by Maldacena and Ludwig 
\cite{ML} and more recently by Ye \cite{Ye}. 
In the bosonic language, the new boundary condition reads: 
$\Phi_s(0^+)=\pi+\Phi_s(0^{-})\,\,,\,\,\Phi_{sf}(0^+)=\pi-\Phi_{sf}(0^{-})$.   
The operator which achieves this change of boundary conditions 
has been constructed 
in the present 
formulation (which is not explicitly $SU(2)$ symmetric) as the product 
$\sigma_z\cdot\cal{U}$  
of the transformation $\cal{U}$ and of 
the twist operator for $\chi^z$. The total dimension of this operator 
is thus $1/16+1/8=3/16$. Within the CFT approach (which is manifestly $SU(2)$ 
invariant) this operator is identified as the primary operator corresponding 
to the fundamental representation $j=1/2$ of the $\widehat{SU_2(2)}_s$ 
Kac-Moody algebra, {\it i.e} to the WZW field $g_{\alpha}$. Note that 
its dimension 
is indeed $j(j+1)/4=3/16$.  
Thus, the explicit solution using 
abelian bosonisation at the Emery-Kivelson solvable point provides us 
with an explicit derivation of the 'fusion principle' of the CFT solution. 

\subsection{Finite-size spectrum}

The whole finite-size spectrum can be simply recovered from the 
Majorana fermion approach (see also \cite{ML}).  
As in the CFT approach, the first task is to construct the spectrum at the 
free-fermion fixed point. 
Let us choose antiperiodic boundary conditions for the 
four original Dirac fermions $\psi_{i\alpha}$. 
The corresponding free-fermion partition function reads:
\begin{equation}
Z_{FF}=\prod_{m=0}^{\infty} (1+q^{m+1/2})^8
\label{zff}
\end{equation}
where $q\equiv e^{-4\pi L/\beta v_F}$. 
We must find how to construct this spectrum in terms of 
$8$ free Majorana fermions, $3$ of them ($\vec{\chi}$) being associated 
with the $\widehat{SU}_2(2)$ spin degrees of freedom, and $5$ being 
associated with the $\widehat{SO_1(5)}$ charge-flavour degrees of freedom. 
It turns out that {\it both} the antiperiodic {\it and} the periodic sector 
must be considered, together with a projection onto states with an even 
total fermion number $F\equiv F_{spin}+F_{c-f}$. The corresponding 
partition function in then obtained in the form (using formulas of Ref.
\cite{PG}):
\begin{eqnarray}
\nonumber 
&Z_{FF}=\mbox{tr}_A {{1+(-1)^F}\over{2}} q^{L_0} 
+ \mbox{tr}_P {{1+(-1)^F}\over{2}} q^{L_0}\\
\nonumber 
&={{1}\over{2}}\prod_{m=0}^{\infty} (1+q^{m+1/2})^8 +
{{1}\over{2}}\prod_{m=0}^{\infty} (1-q^{m+1/2})^8 +\\
\nonumber &8 q^{1/2} \prod_{m=1}^{\infty} (1+q^{m})^8\\
&=q^{1/6}\{{{1}\over{2}} (\sqrt{{{\theta_3}\over{\eta}}})^8 + 
{{1}\over{2}} (\sqrt{{{\theta_4}\over{\eta}}})^8 + 
8 (\sqrt{{{\theta_2}\over{\eta}}})^8 \}
\end{eqnarray}
in which the Jacobi functions $\theta$'s and $\eta$ have argument $q$.
This can be checked to coincide precisely with (\ref{zff}). 

In order to obtain the spectrum at the Kondo fixed point from this 
expression, one must perform a modular transform 
$q\rightarrow w\equiv e^{-\pi \beta v_F/L}$, and  
twist the boundary conditions of the 
three Majorana fermions associated with the spin sector, as explained above. 
This amounts to acting on the states with the operator $(-1)^{F_{spin}}$. 
It turns out that only the sector which was originally antiperiodic at 
the free-fermion fixed-point (and which has been twisted to periodic) 
contributes to the spectrum at the Kondo fixed point.
The resulting partition function (with ground-state energy substracted) 
reads:
\begin{eqnarray}
&Z_{2channel}=
{{1}\over{\sqrt{2}}} w^{-1/48}  \{ (\sqrt{{{\theta_2}\over{\eta}}})^{3} 
(\sqrt{{{\theta_3}\over{\eta}}})^{5} +\\ 
&(\sqrt{{{\theta_3}\over{\eta}}})^{3} 
(\sqrt{{{\theta_2}\over{\eta}}})^{5}\}\\
&=2+4w^{1/8}+10w^{1/2}+12w^{5/8}+26w+\cdots
\end{eqnarray}
in which the $\theta$'s and $\eta$ have argument $w$. 
From this, the lowest energy levels (and degeneracies) are read off as: 
$L\Delta E/\pi v_F= 0(2)\,,1/8(4)\,,1/2(10)\,,5/8(12)\,,1(26),\cdots$.  
This can be checked to coincide with the result obtained from the 
fusion principle (\ref{fusion}) in the CFT approach. 

\subsection{Physical properties at and near the fixed point.} 

Physical properties associated with the 
stable fixed point can also be analyzed in the bosonised language, 
and connections made with the CFT approach. 
It is convenient \cite{EK} to fully 
refermionize the 
hamiltonian in terms of the Dirac fermion $\psi_s$ and its spin-flavour 
counterpart:
\begin{equation} 
\psi_{sf}\equiv {{1}\over{\sqrt{2\pi a_0}}} e^{-i\Phi_{sf}}
\end{equation}
and to represent the impurity spin components in terms of 
three {\it local} Majorana fermions:
\begin{equation}
S^x={{b}\over{\sqrt{2}}}\,\,,\,\,
S^y={{a}\over{\sqrt{2}}}\,\,,\,\,
S^z=-i a\,b
\end{equation}
The hamiltonian can be written as a sum of four terms 
$H=H_c+H_f+H_{sf}+H_s$, where 
$H_c$ and $H_f$ are the free charge and flavour parts, and 
the spin-flavour and spin parts read, in terms of these 
fermionic variables:
\begin{eqnarray}
\nonumber &H_{sf} = i v_F\int_{-\infty}^{+\infty} dx
\psi_{sf}^{\dagger}(x){{\partial}\over{\partial x}} \psi_{sf}(x)\\
&+{{J}\over{\sqrt{\pi a_0}}} [\psi_{sf}^\dagger(0)+\psi_{sf}(0)] b\\
\nonumber &H_s = i v_F\int_{-\infty}^{+\infty} dx
\psi_{s}^{\dagger}(x){{\partial}\over{\partial x}} \psi_{s}(x)\\
&-i\lambda a\,b\, \psi_s^\dagger(0) \psi_s(0)
\end{eqnarray}  
where the coupling constant $\lambda\equiv J_z-2\pi v_F$ 
measures the deviation from the solvable point. The 
interpretation of $H_{sf}$ as a resonant level model is  
very clear in this form. One also sees that only the $b$ 
component of the impurity spin degrees of freedom hybridizes 
at the solvable point, in the same way that only the 
$\psi_{sf}^{+}+\psi_{sf}\propto \chi_z$ component of the 
conduction electron enters $H_{sf}$. $a$ remains a 
local fermion, with non-decaying correlations:
\begin{equation} 
<Ta(0)a(\tau)>= -{{1}\over{2}} \mbox{sign}(\tau)
\end{equation}
This gives rise to  
a residual entropy at the fixed point $S_{imp}=ln2/2$ 
\cite{EK}. In a sense, only half of the impurity and conduction electron 
spin-flavour degrees of freedom hybridise. In contrast to $a$, 
$b$ hybridises with $\chi^z$ and thus acquires dimension 
$1/2$, so that impurity spin correlations at the 
fixed point decay as:
\begin{equation} 
<S(0)S(\tau)>\propto <b(0)b(\tau)> \propto 1/\tau
\end{equation}

Exactly at the solvable point (where one has a quadratic 
fermion hamiltonian), the impurity susceptibility 
$\chi_{imp}$ vanishes and the specific heat coefficient 
$C/T$ is non-singular \cite{EKetc}. The non-Fermi liquid 
singularities in these quantities are controlled by 
the first non-trivial order of perturbation theory in 
the coupling $\lambda$ away from the solvable point. 
Indeed, the corresponding operator $ab\psi^{+}_s(0)\psi_s(0)$, 
of dimension $3/2$, precisely 
coincides with the CFT identification of the leading 
irrelevant operator at the 
non-trivial fixed point. To see this, we first note that 
$b$ can be traded for $\chi^z$ in the expression of this 
operator since the two fields hybridize at the solvable point. 
Then, we note that $\psi^+_s\psi_s \propto \chi^x \chi^y$, 
so that the operator can be written as:
\begin{equation}
i\epsilon_{abc}\chi^a \chi^b \chi^c \propto 
(\vec{\chi}\times \vec{\chi}).\vec{\chi}
\label{irrop}
\end{equation}
The triplet of Majorana fermions $\vec{\chi}$, of dimension $1/2$, 
is the $\vec{\phi}_{j=1}$ primary operator of the Kac-Moody algebra 
$\widehat{SU_2(2)}$, so that (\ref{irrop}) appears as the descendant 
operator $\vec{\cal{J}}_{-1}\cdot\vec{\phi}_{j=1}$, product of 
this operator 
and the spin current, which is precisely the CFT identification. 
Perturbation theory in this operator \cite{AL,EK,EKetc} leads to 
the non-Fermi liquid behaviour $C/T \propto \chi_{imp} 
\propto \mbox{ln}T$ with a universal Wilson ration $R_W=8/3$. 

It is important to note that the  
{\it original} impurity spin variables (before the transformation 
${\cal U}$ is performed) also have a simple representation in terms 
of the Majorana fermions $\vec{\chi}$ and $a,b$. Indeed:
\begin{eqnarray}
\nonumber &S_{+}\equiv S^x+iS^y={\cal U}^{+}\widetilde{S}^{+}{\cal U}
=e^{iab\Phi_s(0)} (b+i\,a) e^{-iab\Phi_s(0)}\\
\nonumber
&= (b+i\,a) \psi_s(0) /\sqrt{2}
\end{eqnarray}
and $S^z=\widetilde{S}^z$. 
Hence, we obtain:
\begin{eqnarray}
\nonumber &S^x = a\,\chi^x(0)-b\,\chi^y(0)\,\,,\,\,
S^y = a\,\chi^y(0)+b\,\chi^x(0)\\
&S^z = i\,a\,b
\end{eqnarray}
The most relevant part of the impurity spin operator, of dimension $1/2$ 
can thus be identified with $a\,\vec{\chi}(0)$ (where the fermion 
$b$ has been traded for $\chi^z(0)$, with which it hybridizes). This 
identification plays a crucial role in the solution of the two-impurity 
problem presented below.

\section{The two-impurity, two-channel Kondo model}
\label{quartet}

We now turn to the 
`Kondo quartet' model, of two spin-1/2 
impurities coupled to two channels of conduction electrons.
This model was first investigated
using numerical renormalization group methods \cite{IJW,KI1,KI2}.
Recently, an analytical solution of the low-energy universal 
properties of the model was found (\cite{AA} and \cite{AA2,JMG,GSLG}).

In its 
original formulation, the problem involves  
two spin-1/2 impurities, 
$\vec{S}_1$, $\vec{S}_2$ located at positions $\pm\vec{R}/2$ and interacting 
with two-channel of free conduction electrons. 
The electron gas is
described by operators $c_{\vec{k}i\alpha}$, where $\vec{k}$ is 
the momentum, $i=1,2$ is the channel index and $\alpha=\uparrow,\downarrow$ 
the spin index. The hamiltonian reads:
\begin{eqnarray}
\nonumber &H\, =\, \sum_{\vec{k}i\alpha} \epsilon(\vec{k}) 
c^+_{\vec{k}i\alpha} c_{\vec{k}i\alpha}
\,+\,J_{K} [\vec{S}_1\cdot\vec{s}\,(+\vec{R}/2)\,+\\
&\vec{S}_2\cdot \vec{s}\,(-\vec{R}/2)]
-I\,\vec{S}_1\cdot\vec{S}_2
\label{origham}
\end{eqnarray}
In these expressions, $\vec{s}(\vec{r})$ is the total conduction electron 
spin density at position $\vec{r}$, $J_K$ is the Kondo coupling  
(which we always take to be antiferromagnetic $J_K>0$) and 
the inter-impurity coupling $I$ has been taken as an independent parameter. 
Of course, such a coupling is generated anyhow from the second-order 
process involving the conduction electron bath ({\it i.e} the RKKY 
interaction, of order $J_K^2$ for small $J_K$). Standard renormalization-group 
arguments allow us to include this coupling by hand (see also 
\cite{ALJ}).

The problem (\ref{origham}) can be considerably simplified by reducing it 
exactly (at low energy) to a one-dimensional model involving 
$8$ left-moving (chiral) 
fermionic fields $\psi_{li\alpha}(x)$, where the extra-index $l=1,2$ 
originates from the presence 
of the two impurity sites. The procedure has been described in detail 
by other authors (see {\it e.g} \cite{ALJ}), and we shall only 
quote the final form of the hamiltonian here:
\begin{eqnarray}
\nonumber &H = i v_F \sum_{l i\alpha} \int_{-\infty}^{+\infty} dx
\psi_{l i\alpha}^{\dagger}(x){{\partial}\over{\partial x}}
\psi_{l i\alpha}(x)\\
\nonumber &+ J_{+} (\vec{S}_1+\vec{S}_2).(\vec{{\cal J}}_1(0)+\vec{{\cal J}}_2(0))\\
\nonumber &+ J_{m} (\vec{S}_1-\vec{S}_2).(\vec{{\cal J}}_1(0)-\vec{{\cal J}}_2(0))\\
\nonumber &+ J_{-} (\vec{S}_1+\vec{S}_2).\sum_{i,\alpha\beta}
(\psi_{1 i\alpha}^{\dagger}(0) {{\vec{\sigma}_{\alpha\beta}}\over{2}}
\psi_{2 i\beta}(0)\\
\nonumber
&+\psi_{2 i\alpha}^{\dagger}(0) {{\vec{\sigma}_{\alpha\beta}}\over{2}}
\psi_{1 i\beta}(0))\\
&- I \vec{S}_1.\vec{S}_2
\label{ham22}
\end{eqnarray}
In this formula 
$\vec{{\cal J}}_l(x) \equiv \sum_{i,\alpha\beta}
\psi_{l i\alpha}^{\dagger}(x) {{\vec{\sigma}_{\alpha\beta}}\over{2}}
\psi_{l i\beta}(x)$ 
denotes the spin-current at position $x$ of conduction electrons corresponding 
to $l=1,2$. The reader is directed to Ref.\cite{ALJ} for the expression 
of the (bare values of) the coupling constants $J_{+},J_{m},J_{-}$ in terms 
of $J_K$ and $R$.
Alternatively, one could work in the even-odd basis:
\begin{equation}
\psi_{ei\alpha}=(\psi_{1,i\alpha}+\psi_{2,i\alpha})/\sqrt{2}\,\,,\,\,
\psi_{oi\alpha}=(\psi_{1,i\alpha}-\psi_{2,i\alpha})/\sqrt{2}
\nonumber
\end{equation}
in terms of 
which the Kondo couplings in Eq.(\ref{ham22}) read:
\begin{eqnarray}
\nonumber 
&(\vec{S}_1+\vec{S}_2)\cdot
(\Gamma_e\vec{{\cal J}}_e(0)+\Gamma_o\vec{{\cal J}}_o(0))\\
\nonumber &+\Gamma_m\,(\vec{S}_1-\vec{S}_2)\cdot
\sum_{i,\alpha\beta}
(\psi_{e i\alpha}^{\dagger}(0) {{\vec{\sigma}_{\alpha\beta}}\over{2}}
\psi_{o i\beta}(0)+\,h.c)
\end{eqnarray} 
Where the couplings $\Gamma_{e,o,m}$ are the ones used in Ref.\cite{IJW}:
\begin{equation}
\Gamma_e=(J_{+}+J_{-})/2\,\,,\,\,\Gamma_o=(J_{+}-J_{-})/2\,\,,\,\,
\Gamma_m=J_m
\end{equation}
The hamiltonian (\ref{ham22}) is invariant under a parity transformation, 
which exchanges the indices 
$l=1,2$ for both impurity spins and conduction electrons: 
\begin{equation}
\mbox{Parity}\,\,:\,\, \psi_{1i\alpha}\leftrightarrow \psi_{2i\alpha}\,\,,\,\,
\vec{S}_1\leftrightarrow \vec{S}_2
\end{equation}
Exchanging the even and odd combinations: 
\begin{equation}
\psi_{e,i\alpha}\leftrightarrow\psi_{o,i\alpha}\,\,;\,\,
\psi_{1,i\alpha}\leftrightarrow\psi_{1,i\alpha}\,\,,\,\,
\psi_{2,i\alpha}\leftrightarrow -\psi_{2,i\alpha}
\end{equation}
is a discrete symmetry of all terms in (\ref{ham22}), except 
the $J_{-}\propto \Gamma_e-\Gamma_o$ coupling. 

When written in the simplified form (\ref{ham22}), the hamiltonian 
is automatically invariant under a particle-hole transformation. 
Breaking particle-hole symmetry in the original hamiltonian 
of the model will induce potential scattering terms at low-energy, 
of the form:
\begin{equation}
V_e\psi^+_e\psi_e\,+\,V_o\psi^+_o\psi_o
\end{equation}

\subsection{Overview of the results}
\label{overview}

We give here a brief overview of the results that will be derived in 
the following. Let us start by discussing the possible fixed points 
of the model in the space of the various coupling constants. There are 
several couplings ($I,J_{\pm},J_m$) in the hamiltonian (\ref{ham22}), 
but we shall show below 
that the whole phase diagram can actually be discussed in terms of 
only {\it two} parameters. These parameters (denoted $x$ and $y$ in 
the following) are phase-shifts which 
depend in a non-universal way on the original coupling constants, and 
can only be calculated analytically in some limiting regime. 
These two phase shifts are the natural variables in terms of which 
the problem is best described, and in terms of which numerical results 
should be interpreted. The precise phase diagram in terms of these 
two parameters will be found in analytical form in Sec.\ref{asymmetric}.
 
Here we follow a route which is physically more transparent, 
and display a {\it schematic version}  
of the phase diagram in the space of two variables, in terms of 
which the phase diagram has been investigated numerically by 
Ingersent and Jones \cite{KI1,KI2}. These two parameters are the 
ratio $I/T_K$ of the RKKY interaction to the Kondo temperature of the 
single impurity model and the ratio 
$(\Gamma_e-\Gamma_o)/((\Gamma_e+\Gamma_o)\,\propto J_{-}$ 
measuring the relative asymmetry 
between the odd and even Kondo couplings. In order to reduce oneself 
to this parameter space, one possibility is to assume that the 
coupling $\delta J\equiv J_{+}-J_{m}$ is taken to be zero 
(we shall see that $\delta J$ actually corresponds to an irrelevant 
operator around the decoupled impurity fixed point). 

\begin{figure}[hbt]
\parbox{\textwidth}{\epsfig{file=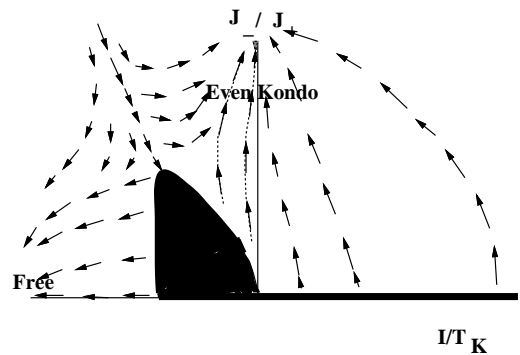,width=7cm,angle=0}} \\
\vspace{9pt}
\caption{Schematic phase diagram in the $(I/T_K,J_{-}/J_{+}\equiv
(\Gamma_e-\Gamma_o)/(\Gamma_e+\Gamma_o))$ plane. The marginal domain 
(made up of a surface and half-line) is indicated in dark. Outside this 
domain, the flow is towards Fermi-liquid fixed points corresponding either 
to even (or odd) Kondo effect or to the free-fermion (=RKKY) fixed point. 
The phase diagram is symmetric under $J_{-}\rightarrow -J_{-}$.}
\label{phasediagsimp}
\end{figure}

In this restricted two-parameter space, the schematic phase diagram 
of the model is depicted in Fig.\ref{phasediagsimp}. The figure is not 
meant to be quantitatively precise (a numerical determination using the 
Wilson RG has been given by Ingersent and Jones in \cite{KI1,KI2}), 
but 
merely indicates the various possible fixed points and directions of 
the RG flow.
Some of the fixed points in Fig.\ref{phasediagsimp} correspond to simple 
limits which can be easily discussed qualitatively \cite{IJW}:

- The fixed point at $I=\delta J=0$, $J_{-}=0$ ($\Gamma_e=\Gamma_o$) 
corresponds to two {\it decoupled} impurities interacting with two 
channels of conduction electrons. The Nozieres-Blandin single impurity 
physics is found at this fixed point. 

- The fixed point at $I=-\infty$, $J_{-}=0$ ($\Gamma_e=\Gamma_o$) 
obviously corresponds to {\it free electrons}. Indeed, for very strong 
antiferromagnetic RKKY coupling, the impurity spins bind into a singlet 
state, leaving a local Fermi liquid with no residual scattering 
for both the even and the odd combinations 
($\delta_e=\delta_o=0$). 

- The fixed point at $I=+\infty$, $J_{-}=0$ ($\Gamma_e=\Gamma_o$) 
corresponds to binding the impurity spins into an $S=1$ triplet state, 
while maintaining perfect symmetry between the odd and even parity 
combinations. As conjectured in \cite{IJW}, it is natural to 
expect that the physics at this fixed point is identical to that of 
a single $S=1$ impurity with {\it twice} two channels of conduction 
electrons, {\it i.e} four channels. Since $4>2\times 1$, this is 
an overscreened non-Fermi liquid fixed point, with 
\cite{AL}: $\chi_{imp}\sim C/T\sim T^{-1/3}$. We shall prove this 
conjecture in the following.

- Finally, the fixed point at $I=0$, $J_{-}\rightarrow +\infty$ 
($\Gamma_e>>\Gamma_o$) has also a simple physical interpretation. 
At this fixed point, the odd electrons decouple and the Kondo effect 
takes place for each impurity with the even combination only. Hence, the 
scattering is characterized by $\delta_e=\pi/2, \delta_o=0$). 
Again, one has a local Fermi-liquid. When the RKKY coupling is ferromagnetic 
($I>0$), we find that an arbitrary small odd-even asymmetry drives 
the system towards this fixed point. In view of the above interpretation 
of the strong ferromagnetic fixed point, we see that this even Kondo 
fixed point must have identical physics to a single-impurity, spin-1, 
2-channel fixed point, which is indeed an exactly screened situation 
and hence a Fermi-liquid. On the antiferromagnetic side, this fixed-point 
controls only part of the parameter space, namely large enough $J_{-}$ and 
small enough $I$ (Fig.\ref{phasediagsimp}). 

Apart from these fixed points that can be found from simple arguments, the 
model also displays a {\it continuous family of 
non-Fermi liquid fixed points} (in dark on Fig.\ref{phasediagsimp}). 
As indicated on Fig.\ref{phasediagsimp}, 
this continuous family is made up of a {\it marginal line} restricted 
to the $J_{-}=0$ ($\Gamma_e=\Gamma_o$) axis when $I>0$ is 
antiferromagnetic, and of a two-dimensional domain 
when $I<0$ is antiferromagnetic. This results from the finding that the 
RKKY coupling is an {\it exactly marginal} perturbation around the decoupled 
impurity fixed point $I=J_{-}=0$, while the even-odd asymmetry is 
{\it relevant} for $I>0$, and again {\it exactly marginal} for $I<0$. 
Within this marginal domain, universal low-energy properties ({\it i.e}  
the low-temperature critical behaviour of the various physical quantities 
and the form of the finite-size spectrum) depend continuously on 
two parameters, according to formulas derived below. 
As an example, for ferromagnetic RKKY and $J_{-}=0$, a power-law behaviour  
$\chi_{imp}\sim C/T \sim 1/T^\theta$ is found, with an exponent varying 
continuously with $I/T_K$ between the limits $\theta(I=0)=0$ 
(corresponding to the Nozieres-Blandin logarithmic behaviour) and 
$\theta=1/3$ (corresponding to the $S=1$, $4$-channel model). 
A separatrix exists in the phase-diagram, which separates the attraction 
basin of the free-electron (RKKY) fixed point and that of the even-parity 
Kondo fixed point. This separatrix hits the boundaries of the marginal 
domain at a multicritical point (Fig.\ref{phasediagsimp}). 

Hence, the main physical message is that inter-impurity effects 
{\it do not} destroy non-Fermi liquid behaviour in this model (in the 
particle-hole symmetric case), but do modify it as compared to the 
single-impurity case and generate a rich variety of new behaviour. The 
competition between the Kondo screening and the RKKY ordering is highly 
involved in this model. This is to be contrasted with the two-impurity 
single-channel model (again at particle-hole symmetry). In this case, 
a RKKY (free electrons) fixed  point controls the physics for sufficiently 
antiferromagnetic  
$I/T_K<(I/T_K)_c$, while a phase with separate Kondo screening of the 
impurities is found for $I/T_K>(I/T_K)_c$. Both of these fixed point are 
Fermi liquids and only the (unstable) multicritical point at $(I/T_K)_c$ 
displays non-Fermi liquid properties.  
   
\subsection{The odd-even symmetric case}
\label{symmetric}

\subsubsection{Symmetries}

We concentrate in this section on the case $J_{-}=0$ 
($\Gamma_e=\Gamma_o$) corresponding to a hamiltonian invariant 
under odd-even exchange. When the coupling between impurities is 
turned off ($I=0, J_{+}=J_m$), the model has a global invariance  
$(SU(2)_{spin}\otimes SO(5)_{fc})^2$. 
At the decoupled impurity {\it fixed point}, 
the symmetry algebra consists in two
copies of a product of
Kac-Moody algebra for spin, channel and charge:
$(\widehat{SU}_2(2)_{s}\otimes SO_1(5)_{fc})^2$  
When coupling the two impurities ($I\neq 0, J_{+}\neq J_m$) 
while keeping $J_{-}=0$, the independent charge and flavour symmetries are 
left unaffected (this is no longer true for $J_{-}\neq 0$), but the spin 
symmetry is reduced to the `diagonal' $SU(2)$ for which $\psi_1$ and 
$\psi_2$ are transformed identically. This reduces 
the conformal symmetry of the spin sector at a fixed point to a 
$\widehat{SU}_4(2)$ algebra. The generators of this algebra are the
sum of the generators of the two $\widehat{SU}_2(2)_s$ for each impurity,
that is the sum of
the spin currents $\vec{{\cal J}}_1(x)+\vec{{\cal J}}_2(x)$.
Hence we must understand how the
product $\widehat{SU}_2(2)_{s}\otimes \widehat{SU}_2(2)_{s}$ can be
decomposed into $\widehat{SU}_4(2)_{s}$ plus some residual degrees of
freedom. This procedure is known in CFT as a `coset construction' \cite{GKO}.
The residual degrees of freedom  
define an algebra $A(2,2)$ such that:
\begin{equation}
\widehat{SU}_2(2)_{s}\otimes \widehat{SU}_2(2)_{s}
=\widehat{SU}_4(2)_{s} \otimes A(2,2)
\label{coset}
\end{equation}
The conservation of the total number of degrees of freedom between the 
two sides of (\ref{coset}) is expressed by the conservation of the total 
central charge. Since the central charge of $\widehat{SU}_2(2)$ is $c=3/2$ 
and that of $\widehat{SU}_4(2)$ is $c=2$, we deduce that the central charge 
of the $A(2,2)$ algebra is $c=3/2+3/2-2\,=1$. 
This algebra is actually known from 
CFT \cite{PG,A22} to be a $N=1$ superconformal unitary model
corresponding to the $m=4$ member of the discrete series with
central charges $c={{3}\over{2}}(1-{{8}\over{m(m+2)}})$. 

\subsubsection{Bosonisation} 
We shall give a schematic derivation of the operator content of the 
algebra $A(2,2)$ using the Majorana fermion representation and abelian 
bosonisation. We introduce two triplets of Majorana fermions 
$\chi_{1,2}^{x,y,z}$ such that the   
$\widehat{SU}_2(2)$ spin currents ${\cal J}_l^a(x)$ ($l=1,2\,$;$a=x,y,z$) 
read: $ {\cal J}_l^a(x)=i \epsilon_{abc} \chi_l^b \chi_l^c$.
These six Majorana fermions can be combined into three Dirac fermions, and 
bosonised as:
\begin{equation}
\chi_1^a(x) + i \chi_2^a(x) =
{{1}\over{\sqrt{2\pi a_0}}}\,e^{-i\Phi_{a}(x)}\,\,;\,\,a=x,y,z
\label{Dirac}
\end{equation}
In terms of these fields, the
total spin current
corresponding to the diagonal $\widehat{SU}_4(2)$ algebra
reads:
\begin{eqnarray}
\nonumber
&{\cal J}^x\equiv {\cal J}_1^x+{\cal J}_2^x=cos(\Phi_y-\Phi_z)\,\,,\,\,
{\cal J}^y=cos(\Phi_x-\Phi_z)\\
\nonumber &{\cal J}^z=cos(\Phi_x-\Phi_y)
\end{eqnarray}
It is convenient to introduce three linear combinations
of boson fields as follows:
\begin{eqnarray}
\nonumber &\Phi = {{1}\over{\sqrt{3}}} (\Phi_x+\Phi_y+\Phi_z)\,\,,\,\,
\mu = {{1}\over{\sqrt{2}}} (\Phi_x-\Phi_y)\\
&\nu = {{1}\over{\sqrt{6}}} (\Phi_x+\Phi_y-2\Phi_z)
\label{bosons}
\end{eqnarray}
Let us also note for further use the inverse relations:
\begin{eqnarray}
\nonumber
&\Phi_x={{1}\over{\sqrt{3}}}\Phi+{{1}\over{\sqrt{2}}}\mu+
{{1}\over{\sqrt{6}}}\nu\,\,,\,\,
\Phi_y={{1}\over{\sqrt{3}}}\Phi-{{1}\over{\sqrt{2}}}\mu+
{{1}\over{\sqrt{6}}}\nu \\
&\Phi_z={{1}\over{\sqrt{3}}}\Phi- 
{{2}\over{\sqrt{6}}}\nu
\label{inverse}
\end{eqnarray} 
In term of these combinations, the components of the total spin current
read:
\begin{eqnarray}
\nonumber &{\cal J}^x=cos({{\mu-\sqrt{3}\nu}\over{\sqrt{2}}})\,\,,\,\,
{\cal J}^y=cos({{\mu+\sqrt{3}\nu}\over{\sqrt{2}}})\\
&{\cal J}^z=cos(\sqrt{2}\mu)
\label{current2imp}
\end{eqnarray}
Note that $\Phi$ does not enter these expressions. 
The two bosons $\mu, \nu$ are associated with the 
$\widehat{SU}_4(2)$ algebra (which has $c=2$), while 
$\Phi$ corresponds to the residual $A(2,2)$ degree of freedom ($c=1$).
This boson is 
compact, with a radius $R=\sqrt{3}$, which means that 
$\Phi$ and $\Phi+2\pi R$ are identified. 
Hence, $A(2,2)$ contains all operators of the form 
$e^{\pm i (n\sqrt{3}+m/2\sqrt{3})\Phi}$ and $\partial^n\Phi$ (with
$n,m$ integers). This yields primary operators of dimensions: 
$(0)\,,\,(1/24)\,,\,(1/6)\,,\,(3/8)\,,\,(1/6+1/2)\,,\,(1)$, 
as summarized in Table \ref{A22op}.
In addition, it can be shown that $\Phi$ and $-\Phi$ must be identified, 
so that $A(2,2)$ is in fact an {\it orbifold theory}. This implies that 
the algebra contains, in addition to the operators above, two twist operators 
of dimension $1/16$ and two twist operators of dimension $9/16$. 
The full set of primary operators of the $A(2,2)$
algebra and their boson
representation (when it exists) is given in Table \ref{A22op}.

\begin{table*}[hbt]
\setlength{\tabcolsep}{1.5pc}
\newlength{\digitwidth} \settowidth{\digitwidth}{\rm 0}
\catcode`?=\active \def?{\kern\digitwidth}
\caption{Primary operators of the $A(2,2)$ algebra. $(\Delta)$ labels an
operator of dimension $\Delta$. The second line displays the bosonic
representation of the operator (when it exists). $NS$ and $R$ stand for
the Neveu-Schwarz and Ramond sectors of the algebra.}
\label{A22op}
\begin{math}
\begin{array}{|c|c|c|c|c|c|c|c|c|c|}\hline
 & & & & & & & & & \\
(0) & ({{1}\over{24}}) & ({1\over16}) & ({1\over 16}) &
({1\over 6}) & ({3\over 8}) & ({1\over 16}+{1\over 2}) &
({9\over 16}) & ({1\over 6}+{1\over 2}) & (1)\\[.1in] \hline
 & & & & & & & & &\\
1 & e^{\pm i {{\Phi}\over{2\sqrt{3}}}} &
 & &
e^{\pm i {{\Phi}\over{\sqrt{3}}}} &
e^{\pm i {\sqrt{3}\Phi\over2}} &
 & &
e^{\pm i {{2\Phi}\over{\sqrt{3}}}} &
\partial\Phi \\[.1in] \hline
NS & R & NS & R & NS & R &  NS & R & NS & NS \\ \hline
\end{array}
\end{math}
\end{table*}
At this stage, we can compare the coset construction made above with 
that encountered in the 2-impurity, {\it single channel} Kondo model 
\cite{AL2}. There, the symmetry algebra at the decoupled impurity 
fixed point is $\widehat{SU_1(2)}\otimes\widehat{SU_1(2)}$. It is broken 
down to $\widehat{SU_2(2)}$ when coupling the two impurities. The coset 
algebra for the residual degrees of freedom thus have central charge 
$c=1+1-3/2\,=1/2$ and identifies with the Ising model:
\begin{equation}
\widehat{SU_1(2)}\otimes\widehat{SU_1(2)}\,=
\,\widehat{SU_2(2)}\otimes\,(Ising)
\end{equation}
The Ising algebra contains three primary operators, of dimension 
$0$ (identity), $1/16$ (order parameter), and $1/2$ (energy density). 
Affleck and Ludwig \cite{AL2} have shown that the properties 
of the non-Fermi liquid 
unstable critical point \cite{JVW} encountered for this model are obtained 
from the decoupled impurity fixed point, by a fusion involving the 
order-parameter operator of the Ising sector. 
In the present case, we shall see that no direct fusion principle 
with an operator of the $A(2,2)$ coset exists. We shall nevertheless 
derive a generalization of this principle involving boundary-condition 
changing operators which {\it are not} in general primary operators 
of the coset algebra, except at special points in the phase diagram. 

\subsubsection{The RKKY interaction as an exactly marginal perturbation} 

We consider the effect of turning on the couplings 
$I$ and $\delta J\equiv J_{+}-J_m$ away from the decoupled impurity 
fixed point (but keeping $J_{-}=0$). At the decoupled point, the impurity 
spin has dimension $1/2$, and the conduction electron spin current has 
dimension $1$. Thus, the RKKY coupling has dimension $1$ and is marginal 
at lowest order, while $\delta J$ has dimension $3/2$ and is irrelevant. 
We shall see that $I$ is actually {\it exactly marginal to all orders}, 
while the dimension of $\delta J$ is actually continuously modified 
as one departs from 
the decoupled fixed point. In order to represent the RKKY term, 
we make use of the two sets of Majorana fermions $\vec{\chi}_1$, 
$\vec{\chi}_2$ above and introduce, as in Sec.\ref{2channel}, two sets of 
Majorana fermions $a_l,b_l$ ($l=1,2$) in order to represent the impurity 
spins $\vec{S}_l$ as in Eq.\ref{majocurrent}. As explained at 
the end of section \ref{2channel}, 
the operator of lowest dimension ($=1/2$) contained in $S_l$ reads 
$a_l\vec{\chi}_l$ in terms of these variables. Hence, the perturbing term of 
lower dimension ($=1$) associated with the RKKY interaction reads:
\begin{equation}
\int dt a_1 a_2 \vec{\chi_1}.\vec{\chi_2}
\end{equation}
In the bosonic language above,
this translates into an induced boundary term in the
$A(2,2)$ sector of the hamiltonian:
\begin{equation}
H_{A(2,2)}={{v_F}\over{4\pi}} \int dx 
({{\partial \Phi}\over{\partial x}})^2
+\tilde{I} (d^{+}d-{{1}\over{2}}) {{\partial \Phi}\over{\partial x}}(0)
\label{hamphi}
\end{equation}
where we have set:
\begin{equation}
d^{+}\equiv (a_1+ia_2)/\sqrt{2}
\end{equation}
with a normalisation $a_1^2=a_2^2=1/2$, so that $\{d,d^+\}=1$. 
Here, $\tilde{I}$ is 
some (non-universal) function of $I$ and $\delta J\equiv J_{+}-J_{m}$,  
which has a perturbative expansion 
$\tilde{I}=I + O(I^2,\delta J^2)$ around the decoupled point. 
Note that the perturbing 
term in Eq.(\ref{hamphi}) is perfectly compatible with the orbifold 
nature of the $A(2,2)$ algebra, provided the transformation
$\Phi\rightarrow -\Phi$ is always made simultaneously with 
$d\rightarrow d^{+}$ on the local degree of freedom.
Since 
$d$ remains a local fermion (with non-decaying correlations), it is 
clear from Eq.(\ref{hamphi}) that the RKKY coupling is associated
with a dimension $1$ operator and is thus 
an {\it exactly marginal perturbation}. 
This results in  
the line of fixed points on the $\Gamma_e=\Gamma_o$ axis 
of Fig.\ref{phasediagsimp}. 

The hamiltonian (\ref{hamphi}) is very similar to the X-ray edge problem 
in the bosonised form \cite{ScSc}. The interaction term on the boundary can 
be absorbed into a redefinition of the field: 
\begin{equation}
\Phi(x) \rightarrow \Phi(x)+\tilde{I}{{\pi}\over{v_F}}(d^+d-{{1}\over{2}})
\,sign(x)
\end{equation}
This changes the boundary condition on the field $\Phi$ by a phase-shift 
$\delta$, such that $\Phi(0^+)-\Phi(0^-)=\pm 2\delta$ when $d^+d-1/2=\pm 1$, 
respectively. The weak-coupling expansion of this phase shift is:
\begin{equation}
{{\delta}\over{\pi}}={{\tilde{I}}\over{2v_F}}=
{{I}\over{2v_F}} + O(I^2,\delta J^2)
\end{equation}
As observed by Schotte and Schotte \cite{ScSc} and also recently discussed 
in \cite{ALX}, this means that the interacting 
hamiltonian is related to the non-interacting one  
by a canonical transformation: 
\begin{equation}
H_{A(2,2)}=U_{\delta}^+ H_{\tilde{I}=0} U_{\delta} \,\,,\,\,
U_{\delta} \equiv e^{i{{2\delta}\over{\pi}}(d^+d-{{1}\over{2}})\Phi(0)}
\end{equation}
This transformation changes the dimension of the operators  
$e^{ig\Phi}$ from ${{g^2}\over{2}}$ to 
${{1}\over{2}}(g\pm {\delta\over\pi})^2$. Note that these dimensions 
do not correspond in general to dimensions of existing operators in 
the $A(2,2)$ algebra. 
This reflects the fact that $U_{\delta}$ is itself not a primary 
operator of this algebra in general. 
Since the available dimensions in this algebra 
are given by $g=n\sqrt{3}+m/2\sqrt{3}$, the maximal 
possible scattering (corresponding to a shift of unity of the integer $m$) 
is reached for ${{\delta_{max}}\over{\pi}}= {{1}\over{2\sqrt{3}}}$. 
Each fixed-point along the marginal line 
in fig.\ref{phasediagsimp} corresponds to a specific value of the phase-shift 
$-\delta_{max}\leq \delta \leq \delta_{max}$, which specifies completely 
all universal properties at the fixed point. 
It is convenient to make use of the normalised parameter:
\begin{equation}
x \equiv {{\delta}\over{\delta_{max}}} = {{2\sqrt{3}\,\delta}\over{\pi}}
\end{equation}
Note that $x>0$ (resp. $x<0$) corresponds to a ferromagnetic 
(resp. antiferromagnetic) RKKY coupling. 
As will be shown below, the marginal lines terminates at infinitely 
strong RKKY coupling $I=+\infty$ on the ferromagnetic side, 
corresponding to maximal scattering $x=+1$, while the marginal behaviour 
only persists for $x>x=x_{min}=-(3-\sqrt{6})\simeq -0.55$ 
on the antiferromagnetic side.  

\subsubsection{Physical properties}

We can investigate the operator content and low-temperature critical 
behaviour of physical quantities 
for non-zero $I$, $\delta J$ (corresponding to a specific value of $x$), 
using the fact that  
a given operator ${\cal O}$ is changed to 
$U_{\delta}{\cal O}U_{\delta}^{+}$ for a non-zero $x$. 
We shall deal first with impurity spin correlation functions. As already 
noted, the most relevant operator (of dimension $1/2$) 
corresponding to each impurity spin 
at the decoupled impurity fixed point reads   
$\vec{S}_l$ is $a_l\vec{\chi}_l$, and we thus have the identification:
\begin{eqnarray}
\nonumber &S_1^a+S_2^a \propto d\,e^{i\Phi_a}+d^{+}e^{-i\Phi_a}\\
&S_1^a-S_2^a \propto d\,e^{-i\Phi_a}+d^{+}e^{i\Phi_a}\,\,,\,\,\,a=x,y,z
\label{spinrep}
\end{eqnarray}
We act on these operators with the transformation $U_{\delta}$, using 
$U_{\delta}\,d\,U_{\delta}^{+}=\exp{(-ix\Phi(0)/\sqrt{3})}\,d$. 
Hence the dimensions $\Delta_{\pm}^{spin}$ of $\vec{S}_1\pm \vec{S}_2$ read, 
for arbitrary $x$:
\begin{equation}
\Delta_{+}^{spin}=
{{1}\over{2}}({{1-x}\over{\sqrt{3}}})^2+{{1}\over{3}}\,\,\,,\,\,\,
\Delta_{-}^{spin}={{1}\over{2}}({{1+x}\over{\sqrt{3}}})^2+{{1}\over{3}}
\end{equation}
From this, we find the low-temperature behaviour of the impurity spin 
susceptibility:
\begin{eqnarray}
\nonumber &\chi_{imp} 
\propto \int^{1/T} {{dt}\over{t^{2\Delta_{+}^{spin}}}} \sim 
{{1}\over{T^{\theta(x)}}}\\
&\theta(x)=1-2\Delta_{+}^{spin}
={{x(2-x)}\over{3}}
\,\,\, (x\geq 0)
\label{chi}
\end{eqnarray}
while $\chi_{imp}$ remains finite on the antiferromagnetic side $x<0$. 
Note that the critical behaviour (\ref{chi}) interpolates continuously 
between the known limits \cite{AL}: 
$\chi_{imp}\sim ln(T)$ at the decoupled impurity fixed point, and 
$\chi_{imp}\sim 1/T^{1/3}$ of the $S=1$, four channel model at 
the strong coupling ferromagnetic fixed point. Similarly, we find that the 
staggered susceptibility $\chi_{st}$ (defined as the response to a field 
coupled to $\vec{S}_1-\vec{S}_2$) is finite on the ferromagnetic side 
$x>0$, and diverges on the antiferromagnetic side as:
\begin{equation}
\chi_{st} \sim
{{1}\over{T^{\theta(-x)}}}\,\,,\,\,\theta(-x)=1-2\Delta_{-}^{spin}=
{{|x|(2+x)}\over{3}} 
\end{equation}

This continous dependance of the critical exponents on $x$ establish 
the existence of a line of fixed points extending on both sides of the 
decoupled impurity fixed point. In order to find the precise extension of 
this line, and the low-temperature behaviour of the specific heat 
along it, we look for the 
leading irrelevant perturbation(s) compatible with 
parity and the even/odd symmetry. At the decoupled impurity fixed point, 
there are two such operators, of dimension $3/2$,  
corresponding to the couplings $J_{+}$ and $J_m$:
\begin{eqnarray}
\nonumber &{\cal O}_{+} \equiv 
(\vec{S}_1+\vec{S}_2).(\vec{{\cal J}}_1(0)+\vec{{\cal J}}_2(0))\\
&{\cal O}_{m} \equiv
(\vec{S}_1-\vec{S}_2).(\vec{{\cal J}}_1(0)-\vec{{\cal J}}_2(0))
\end{eqnarray}
We can find the bosonised form of ${\cal O}_{+}$, using the 
representation of $\vec{S}_1+\vec{S}_2$ given above , and the expression 
(\ref{current2imp}) of the total spin current:
\begin{equation}
{\cal O}_{+} \propto 
d\, e^{i({{1}\over{\sqrt{3}}}\Phi+{{4}\over{\sqrt{6}}}\nu)}\,+\,
d\, e^{i({{1}\over{\sqrt{3}}}\Phi+\sqrt{2}\mu-{{2}\over{\sqrt{6}}}\nu)}
\,+h.c
\end{equation}
Acting with $U_{\delta}$ on this expression, we find that both terms 
give rise for arbitrary $x$ to an operator of dimension:
\begin{equation}
\Delta_{+}={{4}\over{3}}+{{1}\over{2}}({{1-x}\over{\sqrt{3}}})^2 =
1+\Delta_{+}^{spin}
\end{equation}
Similarly, for ${\cal O}_{m}$, we used the bosonised expressions: 
${\cal J}_1^x-{\cal J}_2^x \propto cos(\Phi_y+\Phi_z)$ (and permutations) 
to obtain:
\begin{equation}
{\cal O}_{m} \propto 
d\,e^{-i\sqrt{3}\Phi}\,+\,
d\,e^{i({{1}\over{\sqrt{3}}}\Phi-\sqrt{2}\mu-{{2}\over{\sqrt{6}}}\nu)} 
\,+\,h.c
\label{repom}
\end{equation}
Upon acting with $U_{\delta}$, the first operator becomes of dimension 
$(\sqrt{3}+x/\sqrt{3})^2/2$, while the second has the same dimension than 
${\cal O}_{+}$. Hence the operator of lowest dimension which is generated 
has:
\begin{equation}
\Delta_m = min\,\{{{1}\over{2}}(\sqrt{3}+{{x}\over{\sqrt{3}}})^2\,,
\,\Delta_{+}\}
\end{equation}
The low-temperature behaviour of the specific heat is related 
to the dimension $\Delta$ of the irrelevant operator of lowest 
dimension by:
\begin{equation}
C/T \propto {{\partial^2 F}\over{\partial T^2}} \,\,\,,\,\,
F \propto \int^{1/T} {{dt}\over{t^{2\Delta}}} 
\sim T^{2\Delta-1}
\end{equation}
Hence:
\begin{equation}
C/T \sim T^{-\alpha}\,\,\,, \alpha=3-2\Delta
\end{equation}
For $x>0$ (ferromagnetic side), ${\cal O}_{+}$ has the lowest dimension, 
and we find $\alpha=3-2\Delta_{+}=1-2\Delta_{+}^{spin}=\theta(x)$, 
so that:
\begin{equation}
x>0\,:\,\,\,\, C/T\,\sim\,\chi_{imp}\,\sim T^{-\theta(x)}
\end{equation}
and we expect a universal ($x$-dependent) Wilson ratio, which can be 
calculated using perturbation theory along the lines of Ref.\cite{EKetc}. 
For $x<0$ however (antiferromagnetic side), 
${\cal O}_{m}$ has the lowest dimension, and we find: 
\begin{equation}
x<0\,:\,\,\,\, C/T \sim T^{-\alpha(x)}\,\,,
\alpha(x)=3-2\Delta_m=|x|{{6+x}\over{3}}
\end{equation}
Note that $\chi_{st}$ behaves with a {\it different} exponent. 
The limiting value of $x=x_{min}$, associated with the termination of the 
marginal line for antiferromagnetic coupling, is reached when 
${\cal O}_{m}$ becomes relevant, {\it i.e} $\Delta_m=1$.  
This yields:
\begin{equation}
x_{min}=-(3-\sqrt{6}) \simeq -0.55
\end{equation}
For $x<x_{min}$, the system flows to the `strong antiferromagnetic' 
(or `RKKY') fixed point, and the two impurities bind into a singlet state. 
Obviously, this fixed point has the properties of a local Fermi-liquid.  
On the ferromagnetic side however, the marginal line extends all the way 
up to the infinitely strong RKKY coupling $I=+\infty$ (corresponding 
to maximal scattering $x=+1$), since ${\cal O}_{+}$ is still irrelevant 
at this point. Note that the critical properties derived above coincide, 
for $x=+1$, with that of the spin-1, 4-channel Kondo problem (with spin 
dimension $1/3$ and leading irrelevant operator of dimension $4/3$), 
in agreement with the conjecture made in Ref.\cite{IJW} and with the 
physical picture that the two impurities bind into a $S=1$ triplet 
state with effectively $4$ channel of conduction electrons (because 
of even-odd symmetry).

\subsubsection{Finite-size spectrum}.

We now investigate the finite-spectrum of the model at a given fixed 
point along the marginal line, as a function of $x$. In order to derive the 
spectrum, one first classifies the states 
of the {\it decoupled impurity} fixed point according to the 
$\widehat{SU_4(2)}_s\otimes A(2,2) \otimes 
(\widehat{SO_1(5)}_{fc})^2$ decomposition, 
and then act on 
each state with the transformation $U_{\delta}$. 
This modifies the contribution of the $A(2,2)$ sector to the total energy
of the state.
The dimension $1/16$ of the twist operators can be shown 
to be unchanged by the action of $U_{\delta}$.
Under multiplication by $U_{\delta}$,
the dimension of an operator $e^{ik\Phi}$ is changed to
${{1}\over{2}}(k\pm {\delta\over\pi})^2$, for $d^+d-1/2=\pm 1/2$,
respectively. Hence, we also need to associate to each state an
eigenvalue of $d^+d-1/2=\pm 1/2$ to decide which of the two possible new
dimensions is produced.
This can be done, when constructing the spectrum at the decoupled
point, by keeping track of the relative sign between the
{\it impurity spin} and the {\it total spin} of the state. More
precisely, the impurity spin ($\vec{S}$) is proportional 
to the adjoint operator
of $\widehat{SU_2(2)}$ ($\vec{\chi}$) 
up to a sign which depends on the state. The
product of these two signs for $l=1,2$ yields the eigenvalue of
$2d^+d-1$.
In particular, this `selection rule' is essential to insure that
the spectrum of the
$S=1$ four-channel model is indeed obtained at $x=+1$. This observation is 
also a way to recover this selection rule in an empirical manner. 
 
The resulting spectrum is displayed in Table \ref{spec22}.
\begin{table*}[hbt]
\setlength{\tabcolsep}{1.5pc}
\catcode`?=\active \def?{\kern\digitwidth}
\caption{Finite-size spectrum of the first 76 low-lying states in the 
even-odd symmetric case $J_{-}=0$ ($\Gamma_e=\Gamma_o$).
$j$ is the total spin quantum
number associated with $\widehat{SU}_2(4)^{spin}$.
The second column gives the charge-flavour representation in the
$\widehat{SO}_1(5)^{l=1}\otimes\widehat{SO}_1(5)^{l=2}$ decomposition.
$\sigma$ is the sign of $d^+d-1/2=\pm 1/2$.
The third column displays the $A(2,2)$ operator associated with each
eigenstate at the decoupled impurities fixed point ($I=0$, i.e $x=0$),
whereas the fifth
column displays the corresponding operator at the strong ferromagnetic
fixed point ($I=+\infty$ i.e $x=+1$). The degeneracy of each state is displayed
in the last column, while $\Delta_{tot}$ is the total conformal dimension
at arbitrary $x$. The normalised excitation energies are given by
$L\Delta E/\pi v_F=\Delta_{tot}-\Delta_{tot}^{gs}$.}
\label{spec22}
\begin{math}
\begin{array}{|c|c|c|c|c|c|c|}\hline
j & SO(5)^{l=1}\otimes SO(5)^{l=2} & A(2,2) & \sigma & \Delta_{tot}& A(2,2) &
Deg. \\
 & & decoupled & & & strong ferro. &
\\[.1in]\hline\hline
0 & (\underline{1},\underline{1}) & ({{3}\over{8}}) & +
& {{3}\over{8}} (1+{{x}\over{3}})^2 &
({1\over 6}+{1\over 2}) & 1
\\[.1in] \hline
1 & (\underline{1},\underline{1}) & ({{1}\over{24}})
& - & {{1}\over{3}}+{{(1-x)^2}\over{24}}
& (0) & 3
\\[.1in] \hline
{{1}\over{2}} & (\underline{1},\underline{4})\oplus
(\underline{4},\underline{1}) & ({{1}\over{16}}) & &
{{1}\over{2}} & ({{1}\over{16}}) & 16
\\[.1in]\hline
0 & (\underline{4},\underline{4}) & (0) & + &
{{5}\over{8}}+{{x^2}\over{24}} & ({{1}\over{24}}) & 16
\\[.1in]\hline
0 & (\underline{1},\underline{5})\oplus
(\underline{5},\underline{1}) &
({{3}\over{8}}) & - & {{1}\over{2}}+{{3}\over{8}}
 (1-{{x}\over{3}})^2 & ({1\over 6}) & 10
\\[.1in]\hline
1 & (\underline{1},\underline{5})\oplus
(\underline{5},\underline{1}) & ({{1}\over{24}}) & + &
{{5}\over{6}}+{{(1+x)^2}\over{24}}
& ({1\over 6}) & 30
\\[.1in]\hline
\end{array}
\end{math}
\end{table*}
The normalised excitation energy of a given state,
${{L\Delta E} \over {\pi v_F}}$
(with $L$ the radial length of the bulk system and $v_F$ the Fermi velocity),
is obtained from the total dimension $\Delta_{tot}$ given in the table, 
by the formula:
\begin{eqnarray}
{{L\,\Delta E} \over {\pi v_F}} = \Delta_{tot} - {1 \over 3}
-{(1-x)^2 \over 24} \,\,,\,\, (for\,\, x>0)\\
{{L\,\Delta E} \over {\pi v_F}} = \Delta_{tot} -
{3\over 8}(1+{x\over 3})^2\,\,,\,\, (for\,\,
x<0)
\label{energy}
\end{eqnarray}
The ground-state is the triplet of lowest energy for $x>0$
(ferromagnetic coupling), and
the singlet of lowest energy for $x<0$ (antiferromagnetic coupling).
Comparison with the  
numerical renormalization group 
results of K.Ingersent and B.Jones \cite{IJW,KI1,KI2} reveals that the 
numerical spectra 
are excellently fitted by formula (\ref{energy}) depending on the single 
parameter $x$ (the agreement with the quantum numbers and degeneracies of 
each state has also been checked). Note that, despite the different 
degeneracies of the finite-size ground-state for $x>0$ and 
$x<0$, the residual entropy defined by taking the infinite-volume limit 
first is constant along the marginal line:
\begin{equation}
S_{imp} \equiv \lim_{T\rightarrow 0} \lim_{L\rightarrow\infty} \Delta S 
\,=\,ln 2
\end{equation}
while $S_{imp}$=0 at the strong antiferromagnetic fixed point. 
This is expected from the boundary version of the 'c-theorem'.   

The strong-coupling ferromagnetic fixed 
point found for $x=+1$ deserves special comments. There, the impurity spins 
bind into an $S=1$ triplet state. Thus, it should be possible to describe this 
point, in the spirit of the CFT approach \cite{AL}, by a direct 
fusion in the spin $\widehat{SU_4(2)}$ sector with the triplet operator 
$j=1$ (of dimension $j(j+1)/6=1/3$) applied on the {\it free fermion} 
fixed point. One can check that performing this fusion gives 
precisely the spectrum of a {\it single-impurity, four channel} Kondo model.  
Furthermore, one can also check that exactly the same spectrum  
is obtained when    
acting on the spectrum of the {\it decoupled impurity} fixed point with 
the operator $\exp{i(d^+d-1/2)\Phi(0)/\sqrt{3}}$ (corresponding to 
$U_{\delta}$ for $x=+1$), taking into account the selection rule 
discussed above.  
Note that this operator is then a member of the $A(2,2)$ algebra, 
of dimension $1/24$. Accordingly, the $A(2,2)$ operator labelling 
any eigenstate at $x=0$ is changed at $x=+1$ into another operator 
of the $A(2,2)$ algebra. This strong-coupling assignment is also displayed in 
Table \ref{spec22}. It is compatible with the operator product expansion 
of the $({{1}\over{24}})$ operator with any other operator of the 
$A(2,2)$ algebra:
\begin{eqnarray}
\nonumber&({{1}\over{24}})\times({{1}\over{24}})
\rightarrow (0)+(1)+({{1}\over{6}})\\
\nonumber&({{1}\over{24}})\times({{1}\over{16}})_{NS}\rightarrow
({{1}\over{16}})_{R}+({{9}\over{16}})_{R}\\
\nonumber &({{1}\over{24}})\times({{1}\over{16}})_{R}\rightarrow 
({{1}\over{16}})_{NS}+({{1}\over{16}}+{{1}\over{2}})_{NS}\\
\nonumber&({{1}\over{24}})\times({{1}\over{6}})\rightarrow({{3}\over{8}})+
({{1}\over{24}})\\
\nonumber &({{1}\over{24}})\times({{3}\over{8}})\rightarrow
({{1}\over{6}})+({{1}\over{6}}+{{1}\over{2}})\\
\nonumber&({{1}\over{24}})\times({{9}\over{16}})_{R}\rightarrow  
({{1}\over{16}})_{NS}+({{1}\over{16}}+{{1}\over{2}})_{NS}\\
\nonumber &({{1}\over{24}})\times({{1}\over{16}}+{{1}\over{2}})_{NS}\rightarrow
({{1}\over{16}})_{R}+({{9}\over{16}})_{R}
\end{eqnarray}
The selection rule discussed above dictates, for each state, 
which of the possible operators 
appearing in the r.h.s of these rules are {\it actually generated} 
when going from 
the decoupled impurity spectrum to the strong ferromagnetic one. For example, 
the singlet state in Table \ref{spec22} and the multiplet of degeneracy $10$ 
are both associated with the operator $({{3}\over{8}})$ for $x=0$, but are 
assigned different operators ($({{1}\over{6}}+{{1}\over{2}})$ 
for the first one, 
$({{1}\over{6}})$ for the second) at the strong ferromagnetic point. 
   
\subsection{The effect of odd-even asymmetry}
\label{asymmetric}

We now consider the effect of a non-zero value of the coupling $J_{-}$, 
breaking 
the symmetry between even and odd combinations ($\Gamma_e\neq \Gamma_o)$. 
In the presence of this coupling, the independent flavour and charge 
global symmetry $(SO(5))^2$ no longer holds: only identical 
transformations for both impurities are allowed.   
The symmetry group in the flavour and charge sector is thus 
broken down to the $SO(5)$ diagonal subgroup. 
the sum of the flavour and charge generators. Correspondingly, at a fixed 
point, the model has a  
$\widehat{SU_4(2)}_{s}\otimes \widehat{SO_2(5)}_{fc}$ Kac-Moody symmetry, 
and we must consider the following `coset construction' 
in the flavour-charge sector (in addition to the one above in the spin 
sector):
\begin{equation}
\widehat{SO_1(5)}\otimes \widehat{SO_1(5)} = 
\widehat{SO_2(5)}\otimes K
\end{equation}
It is crucial in this case to take into account the full $SO(5)$ symmetry 
of the charge-flavour sector, and not only its $SU(2)_f\otimes U(1)_c$ 
subgroup. Indeed, this 
excludes some marginal perturbations that would be naively admissible
in the independent flavour/charge language and explains how multiplets
of states at $J_{-}=0$ are broken into multiplets of smaller
degeneracies when $J_{-}\neq 0$.

The current algebra of $\widehat{SO_1(5)}$ is essentially a straightforward 
generalization of that of $\widehat{SU_2(2)}$ to the case of five Majorana 
fermions (it has central charge $c=5/2$). Again, five Dirac fermions are 
built out of the two sets of Majorana's, and converted into five 
bosonic fields $\widetilde{\Phi}_i$. The diagonal combination 
$\widetilde{\Phi}\equiv \sum_i \widetilde{\Phi}_i/\sqrt{5}$ is associated 
with the coset algebra $K$, which has central charge $c=1$. The four 
other combinations correspond to $\widehat{SO_2(5)}$ (with $c=4$). 
The coset $K$ is a subalgebra of the product of cosets 
$A(2,2)_f\otimes \widehat{U(1)}_{c1-c2}$, which have both $c=1$. 
Accordingly, the boson $\widetilde{\Phi}_i$ is a specific linear 
combinations of the two bosons associated with the flavour and charge 
cosets, namely:
\begin{equation}
\widetilde{\Phi} = \sqrt{{{3}\over{5}}}\, \Phi^f + 
\sqrt{{{2}\over{5}}}\,{{\Phi_1^c-\Phi_2^c}\over{\sqrt{2}}}
\label{phitilda}
\end{equation} 

We now ask whether turning on a small coupling $J_{-}$ is a relevant 
perturbation at a given fixed point on the marginal line of  
the previous section. 
To answer this, we consider the corresponding 
operator:
\begin{eqnarray}
\nonumber &{\cal O}_{-} \equiv 
(\vec{S}_1+\vec{S}_2).\sum_{i,\alpha\beta}
(\psi_{1 i\alpha}^{\dagger}(0) {{\vec{\sigma}_{\alpha\beta}}\over{2}}
\psi_{2 i\beta}(0)\\
&+\psi_{2 i\alpha}^{\dagger}(0) {{\vec{\sigma}_{\alpha\beta}}\over{2}}
\psi_{1 i\beta}(0))
\end{eqnarray}
A naive counting of dimension would suggest that this operator has dimension 
$1/2+1=3/2$ at the decoupled impurity fixed point. This is incorrect 
however, since ${\cal O}_{-}$ is a product of an impurity and 
fermion parts which are not independent. Instead, we shall see that 
${\cal O}_{-}$ is, to lowest order in $J_{-}$, a marginal operator 
(of dimension $1$) at the decoupled impurity fixed point. 
It is convenient to find the bosonised form of 
${\cal O}_{-}$ at this fixed point. 
$\vec{S}_1+\vec{S}_2$ is represented as in eq.(\ref{spinrep}), 
and we need to concentrate on 
$\psi_{1 i\alpha}^{\dagger}(0) {{\vec{\sigma}_{\alpha\beta}}\over{2}} 
\psi_{2 i\beta}(0)\,+h.c$. This operator is a spin triplet, and a flavour 
and charge singlet. Using the spin-flavour-charge decomposition 
$\widehat{SU}_2(2)_{s}\otimes \widehat{SU}_2(2)_{f}\otimes
\widehat{U}(1)_{c}$ for each impurity index $l=1,2$, we have to build a 
triplet combination out of two $j=1/2$ operators of the 
$\widehat{SU}_2(2)_{s}$ algebra (each of dimension 
$j(j+1)/4=3/16$, so that the spin contribution to the overall dimension is 
$3/16+3/16=3/8$). Similarly, we build a singlet operator out of 
two $j=1/2$ operators of the 
$\widehat{SU}_2(2)_{f}$ algebra (also of dimension $3/8$), 
and a singlet charge combination (of dimension $1/8+1/8=1/4$). 
If we represent each fermion $\psi_{li\alpha}^{+}$ using  
{\it non-abelian} bosonisation (Eq.(\ref{nonab})), as 
$g_{l\alpha}^{+}h_{li}^{+}e^{i\Phi_l^c/2}$, 
this corresponds to forming the combination 
$g_{1\alpha}^{+}\vec{\sigma}_{\alpha\beta}g_{2\beta}\,
h_{1i}^{+}h_{2i}\,e^{i(\Phi_1-\Phi_2^c)^c/2}$. The flavour singlet combination 
$h_{1i}^{+}h_{2i}$ is built out of the three operators (of dimension $1/8$)  
$e^{i\Phi^f_a/2}, a=x,y,z$ (which give a $\pi/2$ phase shift to one 
of the three flavour Dirac fermions), as 
$\exp{i(\Phi^f_x+\Phi^f_y+\Phi^f_z)/2}=
\exp{i\sqrt{3}\Phi^f/2}$. Similarly, we built the spin triplet 
$g_{1\alpha}^{+}\vec{\sigma}_{\alpha\beta}g_{2\beta}$ 
out of the $e^{i\Phi_a/2}$ operators as: 
$\exp{i(-\Phi_x+\Phi_y+\Phi_z)/2)}$ (and permutations), {\it i.e} 
as: $\exp{i(\sqrt{3}\Phi/2-\Phi_a)}\,\,,a=x,y,z$. Multiplying by 
$\vec{S}_1+\vec{S}_2$ given by Eq.(\ref{spinrep}), we find that the 
operator of lowest dimension contained in ${\cal O}_{-}$ reads 
in bosonised form:
\begin{eqnarray}
\nonumber
&d\,\mbox{exp}\{i({{\sqrt{3}}\over{2}}\Phi(0)+{{\sqrt{3}}\over{2}}\Phi^f(0)\\
&+{{1}\over{\sqrt{2}}}{{\Phi^c_1(0)-\Phi^c_2(0)}\over{\sqrt{2}}})\}\,+h.c
\label{repomin}
\end{eqnarray}
Hence, this operator is of dimension $1$ at the decoupled impurity 
fixed point, as announced. Under the action of $U_{\delta}$, the 
spin part is transformed into $\exp{i(\sqrt{3}/2-x/\sqrt{3})\Phi(0)}$ 
and hence the dimension is changed to:
$\Delta_{-}(x)={{1}\over{2}}({{\sqrt{3}}\over{2}}-{{x}\over{\sqrt{3}}})^2+
{{5}\over{8}}$. Since $\Delta_{-}(x)<1$ for $x>0$, we conclude that the 
coupling $J_{-}$ is {\it a relevant perturbation on the ferromagnetic 
side}, and that the system flows away from the marginal line. 
For e.g $\Gamma_e>\Gamma_o$, the flow must be towards a fixed point where 
the odd combination eventually decouples, and thus corresponds \cite{IJW} 
to an even-parity Kondo effect with $S=1$ and two channel 
of conduction electrons, which 
is a Fermi liquid fixed point (cf. fig.??). 
On the antiferromagnetic side of the line ($x<0$) however, we find that 
$\Delta_{-}(x)>1$ and thus that a small $J_{-}$ is 
{\it an irrelevant perturbation}. Hence, {\it the marginal behaviour must be 
preserved} for small enough $J_{-}$ on this side.  

In order to find the new marginal operators generated by the coupling 
$J_{-}$ for $x<0$, we consider the perturbation expansion in the 
operator (\ref{repomin}). 
This operator is irrelevant, but its operator product 
expansion with itself will generate the  
marginal operator $(d^+d-1/2)\partial_x\Phi(0)$ 
(inducing simply a renormalisation of $\widetilde{I}$, 
proportional to $J_{-}^2$ for small $J_{-}$), but also the marginal 
operator 
$(d^+d-1/2)\partial_x \widetilde{\Phi}(0)$, where $\widetilde{\Phi}$ 
is precisely the linear combination of flavour and charge bosons 
(\ref{phitilda}) corresponding to the coset $K$. This operator was forbidden 
for $J_{-}=0$.  
The fact that precisely this combination appears is due to the $SO(5)$ 
symmetry of the theory. Had we used the 
separate flavour-charge decompositions of, we would have 
concluded incorrectly that {\it two} new marginal perturbations 
$\partial_x\Phi^f(0)$ and $\partial_x(\Phi^c_1-\Phi^c_2)$ were {\it 
a priori} allowed, corresponding to the two $c=1$ cosets 
$A(2,2)_f$ and $\widehat{U(1)}_{c1-c2}$. 
Hence, we find that, in addition to the induced boundary term (\ref{hamphi}) 
in the spin sector, we have a single other boundary term induced in the 
flavour-charge sector, which reads:
\begin{equation}
(d^+d-{{1}\over{2}})\,{{\partial\widetilde{\Phi}}\over{\partial x}}(0)
\label{hamphisf}
\end{equation}
The problem is again solved by a canonical transformation which 
now involves a single additional phase shift $\delta'$ 
associated with (\ref{hamphisf}), 
depending on the 
couplings in a non-universal manner with 
$\delta'\propto J_{-}^2$ for small $J_{-}$. In the following, we make 
use of a parameter $y$ such that $\delta'/\pi=\sqrt{5}y/2$. 
The overall canonical transformation reads:
\begin{equation}
U_{\delta\delta'} =
\exp \{ i(d^+d-{{1}\over{2}})\,({{x}\over{2\sqrt{3}}}\,\Phi(0)+
{{\sqrt{5}y}\over{2}}\,\widetilde{\Phi}(0)) \}
\end{equation}

\begin{figure}[hbt]
\parbox{\textwidth}{\epsfig{file=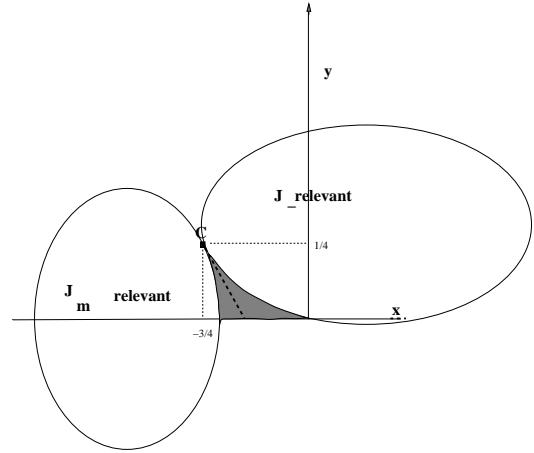,width=7cm,angle=0}} \\
\vspace{9pt}
\caption{The marginal domain in the plane of the phase shift 
parameters $(x,y)$ is made up of the shaded area and the half-axis 
$x>0,y=0$. The (tangent) ellipses inside which the operators 
${\cal O}_{-}$ and ${\cal O}_{m}$ become relevant are indicated: see text 
for a detailed explanation.}
\label{phasediagtrue}
\end{figure}

For antiferromagnetic RKKY coupling ($x<0$), the marginal line extends 
into a two-dimensional marginal domain parametrized by the two parameters 
$x$ and $y$. In order to find its precise extension in the $(x,y)$ plane,    
we discuss the operator content at a given fixed point of the marginal 
domain, concentrating on the irrelevant operator(s) of lowest dimension. 
This analysis will also yield the low-temperature behaviour of physical 
quantities. 
The boundaries of the domain will be characterized by one of 
these operators becoming relevant.
It turns out that we have to discuss only ${\cal O}_m$ and ${\cal O}_{-}$, 
all other operators (including ${\cal O}_{+}$) being of higher dimension in 
this region of the phase diagram. We use the bosonised form (\ref{repomin}) of 
${\cal O}_{-}$ to find the dimension of the transformed operator 
$U_{\delta\delta'}{\cal O}_{-}U_{\delta\delta'}^{+}$, which reads:
\begin{equation}
\Delta_{-}(x,y) = {{1}\over{6}}({{3}\over{2}}-x)^2\,+\,
{{5}\over{2}}({{1}\over{2}}-y)^2
\label{dimomin}
\end{equation}
Similarly, using eq.(\ref{repom}) for ${\cal O}_m$: 
\begin{equation}
\Delta_m(x,y) = {{3}\over{2}}(1+{{x}\over{3}})^2\,+\,
{{5}\over{2}}y^2
\label{dimom}
\end{equation}
Note that these expressions coincide with those established above in the 
limit $y=0$. Thus, we find that these operators are {\it irrelevant} 
($\Delta>1$)  
{\it outside} the ellipses of equations: 
$\Delta_{-}(x,y)=1$ (for ${\cal O}_{-}$) and 
$\Delta_{m}(x,y)=1$ (for ${\cal O}_{m}$). 
The location of these ellipses in the $(x,y)$ plane is depicted on 
fig.\ref{phasediagtrue}.
Remarkably, they are tangent at a single point 
$(x_c,y_c)=(-3/4,1/4)$. The shaded area in fig.\ref{phasediagtrue}, 
delimited by 
these two ellipses and the two coordinate axis, together with 
the whole half-axis $y=0,x\geq 0$ corresponds to the 
marginal domain. When crossing the boundary to the right of the point 
$(x_c,y_c)$ (`ferromagnetic side')
, ${\cal O}_{-}$ becomes relevant and the system flows to 
the even-parity (or odd-parity) Fermi-liquid fixed point (except 
when $y$ is strictly zero). When crossing the boundary to the left 
of $(x_c,y_c)$ (`antiferromagnetic side'), ${\cal O}_{m}$ becomes 
relevant and the system flows to the strong antiferromagnetic Fermi-liquid 
fixed point $I=-\infty$. Note that $y$ is even in $J_{-}$ and 
thus that the phase diagram in the $y>0$ half-plane yields a symmetric 
phase diagram in the coupling $J_{-}$, with odd and even coupling exchanged.  
Thus, we identify $(x_c,y_c)$ as the {\it multicritical 
point} separating the attraction basins of two different fixed points 
when the boundary of the marginal domain is crossed. It is worth 
noticing that at this point the dimension of the operator 
$U_{\delta\delta'}$ becomes $x_c^2/24+5y_c^2/8=1/16$, hence suggesting that 
a direct interpretation of this point as a fusion from the decoupled 
fixed point with some twist operator is probably possible. This 
is reminiscent of the multicritical point of the single-channel, 
two impurity model \cite{AL2}. 

We also discuss the low-temperature behaviour of the impurity 
susceptibilities and specific heat. Acting on the bosonised form 
(\ref{spinrep}) of the operators $\vec{S}_1\pm \vec{S}_2$ with 
$U_{\delta\delta'}$, we find their new dimensions:
\begin{eqnarray}
\nonumber&\Delta_{+}^{spin}={{1}\over{3}}+{{(1-x)^2}\over{6}}+
{{5y^2}\over{2}}\\
\nonumber &\Delta_{-}^{spin}={{1}\over{3}}+{{(1+x)^2}\over{6}}+
{{5y^2}\over{2}}
\end{eqnarray}
Hence, the impurity susceptibility is found to be finite 
within the whole antiferromagnetic 
part $x<0$ of the marginal domain, while the staggered susceptibility 
(defined as above) diverges as:
\begin{equation}
\chi_{st}\sim T^{-\theta}\,\,\,\,,\,\,\,\,
\theta={{|x|(2+x)}\over{3}}-5y^2 
\end{equation}
In order to find the behaviour of the specific heat inside the 
marginal domain, we have to find which of the two irrelevant 
operators ${\cal O}_m$ and ${\cal O}_{-}$ has lowest dimension. 
Using Eqs.(\ref{dimomin},\ref{dimom}), 
we find that $\Delta_{-}<\Delta_m$ to the right of 
the straight line $3x+5y=1$ (also depicted as a dashed line
 on fig.\ref{phasediagtrue}), 
while $\Delta_{m}<\Delta_{-}$ to its left. 
Hence, the specific-heat behaves as:
$C/T \sim T^{-\alpha}$ with:
\begin{eqnarray}
\nonumber&\alpha=3-2\Delta_{-}(x,y)\,\,\,\, for\,\, 3x+5y<1\,\,,\,\,
(y\neq 0)\\
&\alpha=3-2\Delta_{m}(x,y)\,\,\,\, for\,\, 3x+5y>1
\end{eqnarray}
The straight line $3x+5y=1$ 
contains $(x_c,y_c)$ (where $\Delta_{-}=\Delta_m=1$) and crosses the 
$y=0$ axis at $x=-1/3\,>x_{min}$. On this axis however, the operator 
${\cal O}_{-}$ is not allowed because of the odd-even symmetry, and the 
low-temperature behaviour of $C/T$ was found to be controlled by 
${\cal O}_m$. Hence our results lead to the somewhat surprising conclusion 
that the exponent of the specific heat should change discontinuously 
for $-1/3<x<0$ when $y$ is turned on. 

The finite-size spectrum of the model in the marginal surface can be 
obtained in a similar manner than for the even-odd symmetric case, by 
acting, for each state,  on the operator in the coset $A(2,2)\otimes K$ 
with the transformation $U_{\delta \delta'}$.   
The resulting spectrum is also in good agreement with available 
numerical data, and will be reported in another article \cite{AA2,GSLG}. 

\section{Conclusion}
\label{conclusion} 

We have shown that the competition between inter-impurity (RKKY) interactions 
and the (partial) screening associated with the Kondo effect, lead to 
a continuous two-parameter family of new non-Fermi liquid fixed points in 
the two-impurity, two-channel Kondo model at particle-hole symmetry. An 
analytical solution of the universal low-energy properties of this model 
has been obtained, using a combination of bosonisation and conformal field 
theory techniques. 

Possible physical realizations of this model include magnetic impurities 
in coupled Heisenberg spin chains \cite{AA2} and coupled quantum dots 
devices \cite{M,ASZ}. In those cases, one often faces an RKKY interaction 
which is not isotropic in spin space. This can be solved by bosonisation 
along the same 
lines as the isotropic case considered in this paper. The perturbation 
then acts not only in the coset algebra, but also in the 
$\widehat{SU}_4(2)$ spin sector, and additional marginal operators 
(hence additional phase shifts) are induced.      

An outstanding question that has not been dealt with above  
is the effect of particle-hole symmetry breaking on this family of fixed 
points. We have shown \cite{GSLG} that it is a relevant perturbation 
near the strong ferromagnetic fixed-point (and hence, by continuity, along 
the whole marginal line for $x>0$). The situation on the antiferromagnetic 
side is currently under investigation.
\section{Acknowledgments}
We are extremely grateful to K.Ingersent and B.Jones 
for a very fruitful correspondance and for the
communication of their unpublished NRG results \cite{KI2}. We also 
acknowledge discussions with N.Andrei and E.Kiritsis and collaboration 
with A. Ludwig.
The hospitality, at various stages of this work, of the ITP, 
Santa Barbara (under NSF grant 
PHY94-07194), the ICTP (Trieste), the Aspen Center for Physics, 
and the Rutgers University Physics 
Department is gratefully acknowledged.

\end{document}